%% file: main.tex
\documentclass[aps,prl,twocolumn,superscriptaddress,nofootinbib,floatfix,noshowpacs]{revtex4-1}
\pdfoutput=1
\newif\ifcomment
\usepackage{dcolumn}
\usepackage{bm}
\usepackage{amsmath,amssymb,amscd,amsfonts}
\usepackage{color,hyperref,url}
\usepackage{listings}
\usepackage{slashed}
\usepackage{color}
\usepackage[pdftex]{graphicx}
\usepackage{epstopdf}
\usepackage{epsfig}
\usepackage{grffile}
\usepackage{relsize}
\usepackage{multirow}
\usepackage{xspace}
\usepackage{floatrow}
\usepackage{calrsfs}
\usepackage{txfonts}
\graphicspath{{./img/}}
\usepackage{soul}

\newcommand{\beq}{\begin{equation}}
\newcommand{\eeq}{\end{equation}}
\newcommand{\ba}{\begin{array}}
\newcommand{\ea}{\end{array}}
\newcommand{\bea}{\begin{align}}
\newcommand{\eea}{\end{align}}
\newcommand{\bi}{\begin{itemize}}
\newcommand{\ei}{\end{itemize}}
\newcommand{\ben}{\begin{enumerate}}
\newcommand{\een}{\end{enumerate}}
\newcommand{\bc}{\begin{center}}
\newcommand{\ec}{\end{center}}
\newcommand{\bl}{\begin{flushleft}}
\newcommand{\el}{\end{flushleft}}
\newcommand{\br}{\begin{flushright}}
\newcommand{\er}{\end{flushright}}

\renewcommand{\l}{\left}
\renewcommand{\r}{\right}

\let\OLDthebibliography\thebibliography
\renewcommand\thebibliography[1]{
  \OLDthebibliography{#1}
  \setlength{\parskip}{0pt}
  \setlength{\itemsep}{5pt plus 0.3ex}
}
\input{commands.tex}

\begin{document}

\title{Application of radial basis functions neutral networks in spectral functions}

\author{Meng Zhou}
\email{MengZhou@pku.edu.cn}
\affiliation{Department of Physics and State Key Laboratory of Nuclear Physics and Technology, Peking University, Beijing 100871, China}
\affiliation{Collaborative Innovation Center of Quantum Matter, Beijing 100871, China}

\author{Fei Gao}
\email{gao@thphys.uni-heidelberg.de}
\affiliation{Institut f{\"u}r Theoretische Physik,
	Universit{\"a}t Heidelberg, Philosophenweg 16,
	69120 Heidelberg, Germany}

\author{Jingyi Chao}%
\email{jyc.lou@gmail.com}
\affiliation{
College of Physics and Communication Electronics, Jiangxi Normal University, Nanchang, Jiangxi 330022, China}

\author{Yu-Xin Liu}
\email{yxliu@pku.edu.cn}
\affiliation{Department of Physics and State Key Laboratory of Nuclear Physics and Technology, Peking University, Beijing 100871, China}
\affiliation{Collaborative Innovation Center of Quantum Matter, Beijing 100871, China}
\affiliation{Center for High Energy Physics, Peking University, Beijing 100871, China}

\author{Huichao Song}
\email{huichaosong@pku.edu.cn}
\affiliation{Department of Physics and State Key Laboratory of Nuclear Physics and Technology, Peking University, Beijing 100871, China}
\affiliation{Collaborative Innovation Center of Quantum Matter, Beijing 100871, China}
\affiliation{Center for High Energy Physics, Peking University, Beijing 100871, China}

\date{\today}
\begin{abstract}\noindent
The reconstruction of spectral function from correlation function in Euclidean space is a challenging task. In this paper, we employ the Machine Learning techniques in terms of the radial basis functions networks to reconstruct the spectral function from a finite number of correlation data. To test our method, we first generate one type of correlation data using a mock spectral function by mixing several Breit-Wigner propagators. We found that compared with other traditional methods, TSVD, Tikhonov, and MEM,
our approach gives a continuous and unified reconstruction for  both positive definite and negative  spectral function,  which is especially useful for studying the  QCD phase transition. Moreover,  our approach has considerably better performance in the low frequency region.  This has advantages for the extraction of transport coefficients which are related to the zero frequency limit of the spectral function. With the mock data generated through a model spectral function of stress energy tensor, we  find  our method  gives a precise and stable extraction of the transport coefficients.
\end{abstract}



\maketitle
\section{I. Introduction}\label{intro}
The main goals of relativistic heavy ion collisions are to create the quark gluon plasma (QGP) and to study the  QCD phase diagram. The heavy ion programs at Relativistic Heavy Ion Collider (RHIC) and the Large Hadron Collider (LHC) have discovered the strongly coupled QGP which behaves like a nearly perfect liquid with small viscosity ~\cite{Gyulassy:2004zy,Adcox:2004mh,Heinz:2013th,Gale:2013da,Song:2017wtw,Song:2010mg,Schenke:2010rr,Xu:2016hmp,Bernhard:2016tnd,Zhao:2017yhj,Zhu:2016puf,Bernhard:2019bmu,Everett:2020yty}. In the theoretical sides,  the QCD phase diagram and transport properties of the QGP have been extensively studied by ~\cite{Aarts:2002cc,Gupta:2003zh,Nakamura:2004sy,Arnold:2000dr,Arnold:2003zc,Kovtun:2004de,Braby:2009dw,Chakraborty:2010fr,Haas:2013hpa,Ding:2016hua,Qin:2013aaa,Gao:2015kea,Gao:2016qkh,Gao:2016hks,Gao:2017gvf,Gao:2020qsj,Fischer:2018sdj,Isserstedt:2019pgx,Maelger:2019cbk,Ding:2020zpp}. In these contexts, the spectral functions (SPF) of the correlation functions play an essential role to capture the feature of QCD matters in medium and to build the relation between QCD and the experimental data~\cite{Hatsuda:1986gu, Hatsuda:1985eb}.

In the hadronic and QGP phases, the spectral functions of two point correlation function behave quantitatively different since confinement prevents the two point correlation function of quark or gluon from appearing as the mass pole in real time axis~\cite{Asakawa:2000tr,Schafer:1993ra,Shuryak:1993kg,Fischer:2017kbq,
Cyrol:2018xeq,Binosi:2019ecz,
Horak:2021pfr}. As for the extraction of  transport coefficients, the spectral functions of the stress energy tensor correlation function are required  in the description of the Kubo formula.  These transport coefficients can be expressed as  the low-frequency part of the   spectral functions ~\cite{Meyer:2007ic, Meyer:2011gj, Amato:2013naa, Christiansen:2014ypa,
Qin:2014dqa}. However, the required real-time information, ie. spectral function,  is very difficult to be obtained directly through the first-principle calculation, since  most of the non perturbative QCD computations including lattice simulation and functional QCD methods  are carried out in Euclidean space~\cite{Asakawa:2000tr}.
It is known that the SPF is related to the correlation function by an integral manner with respect to the full momentum spaces.
With the limited inputs from Euclidean space, it then becomes an ill-posed problem   where the solution of inverse mapping suffers several undetermined properties such as the existence, uniqueness and stability.
The limited numerical data of correlation functions drive to large uncertainties in extracting SPF, i.e., it produces a family of allowed solutions, but most of them are highly oscillated and not physically meaningful~\cite{hansen_numerical_1992,kirsch2011introduction}.

In order to obtain the physical solution,  many methods have been suggested. For example, a truncated Singular Value Decomposition (TSVD)~\cite{hansen_truncated_1990,hansen_numerical_1992,chen_truncated_2017} is a straightforward method that has been widely employed to analyze the ill-conditioned inverse problem. By treating the high frequency part as noise and then truncating it in SVD method, the approximate solution converges to the exact solution.
The regulating term that makes the spectral function less oscillating  is another choice as in  Tikhonov method~\cite{groetsch1984theory,Dudal:2013yva}.
The most commonly used and powerful reconstructing approach of the spectral function is the maximum entropy method (MEM)~\cite{Asakawa:2000tr,Ikeda:2016czj,Gao:2014rqa,Qin:2014dqa,Gao:2016jka}, where the uniqueness of the extracted spectral representation can be achieved through introducing a default model for prior conditions.

Recently, deep neural networks have been employed to solve the problem of SPFs in supervised learning~\cite{yoon2018analytic, fournier2020artificial, Kades:2019wtd}. The previous computations applied a direct mapping from correlations to the SPFs with a long time training on the prepared datasets. The way of supervised learning  could quickly get highly accurate predictions as long as the test samples belong to the same distribution of the training set. However,  it might break down without warning if the test samples are from different domains' datasets. Stimulating by the recent development of neural networks, we adopted the radial basis functions network (RBFN)~\cite{broomhead1988radial,schwenker2001three}  which is a multilayered perceptron model that is widely used in classification, regression, feature extraction, etc~\cite{beheim2004new,wang2002point,carr1997surface,chen2018deep}.  The main strategy of this approach is to transform the inverse mapping problem into
calculating  the linear weights of the radial basis functions (RBF), which enables a smooth and continuous reconstruction which hasn't been accomplished by other methods. Besides, the programme runs fast with the adopted  matrix method used in this paper.

This  paper is organized as  the following. Sec.~II introduces the notorious problem of spectral reconstruction  and briefly reviews several state-of-the-art methods widely used to reconstruct the spectral function. Sec.~III describes our RBFN method. Section IV shows the numerical results from our RBFN method, together with a comparison to the results from other traditional methods described in Sec.~II.  Sec.~V summaries this paper and discusses possible future works.

\section{II. Spectral reconstruction and existing methods}\label{sec:SPF}
\subsection{A. Spectral reconstruction as an ill-posed problem}
 The correlation functions  in Euclidean space can be calculated via the first-principle non perturbative approaches~\cite{Meyer:2011gj,Qin:2014dqa,Ding:2015ona}.  The SPF $\rho(\omega, T)$ is related to the correlation function $G(\tau, T)$ through an integral spectral representation:
\begin{equation}
  G\l(\tau,T\r)=\int^\infty_0 \frac{d\omega}{2\pi}\rho\l(\omega ,T\r)K\l(\omega,\tau,T\r),\label{eq:maineq}
\end{equation}
where the integration kernel is
\begin{equation}\label{eqn_kernel}
  K(\omega,\tau,T)=\frac{\text{cosh}\l(\omega\tau-\frac{\omega}{2T}\r)}{\text{sinh}\l(\frac{\omega}{2T}\r)}.
\end{equation}
$\tau$ is the imaginary time and $T$ stands for temperature. Above equation  belongs to the Fredholm integral equation of the first kind, which  is ill-posed since there are many solutions to a certain set of input data.

The spectral representation has been studied in several cases. Firstly, the spectral function of the propagator reveals the dispersion relation of the respective particle. As known by the Kallen-Lehmann spectral representation, it
gives a general expression for the two point Green function of quantum field theory in vacuum, which is written as:
\begin{equation}
\label{eq:spectral}
G(p^2)=\frac{1}{2\pi}\int_0^{\infty} d\mu^2\rho(\mu^2)\frac{1}{p^2+\mu^2},
\end{equation}
with $\rho(p^2)$ being the spectral function of the propagator.  Similar forms have been generalized into finite temperature and chemical potential regions where the information of the spectral function are related to the chiral and deconfinement phase transitions directly \cite{Rapp:2009yu, Bashir:2012fs, Tripolt:2016cey}.

Secondly, another important goal is to obtain the transport coefficients such as the shear viscosity $\eta$, which could be derived from the spatial traceless part $\pi_{ij}$ of stress-energy-momentum tensor correlation function via Kubo relation~\cite{Christiansen:2014ypa}:
\begin{align}\label{eqn:eta}
  \eta = \lim_{\omega\rightarrow 0}\frac{1}{20}\frac{\rho_{\pi\pi}(\omega,\vec{0})}{\omega},
\end{align}
where
\begin{equation}
\rho_{\pi\pi}(\omega,\vec{p})=\int\frac{d^4 x}{(2\pi)^4}e^{-i\omega x_0 + i\vec{p}\vec{x}}\left \langle[\pi_{ij}(x), \pi_{ij}(0)]\right\rangle.
\end{equation}
The above correlation function is defined in Minkowski space and is deformed to spectral representations via analytic continuation where
\begin{equation}
G(i\omega_n,\mathbf{k}) = \int \frac{d\omega'}{2\pi}\frac{\rho_{\pi\pi}(\omega',\mathbf{k})}{\omega'-i\omega_n}.
\end{equation}
The Matsubara frequency $\omega_n=i(2n+s)\,\pi T$ in imaginary time formula with fermionic ($s=1$) or bosonic ($s=0$). And its Fourier transformation  defined as
\begin{equation}
G(\tau,\mathbf{k}) \equiv T \sum_n e^{-i\omega_n \tau} G(i\omega_n, \mathbf{k}),
\end{equation}
which is   the standard Euclidean quantity as in  Eq.(\ref{eq:maineq}) applied in this ill-posed problem.
The extraction of transport coefficients requires the giving  spectral function to be  smooth and stable at the low frequency limit, which will then show to be the advantage of the method developed here.

As mentioned above, up to now, the lattice simulations and other functional methods are usually used to compute the correlation functions at a finite set of discrete points in the Euclidean space.  Following the obtained numerical data, the integral equation is usually discretized as:
\begin{equation}\label{eq:MatFormOfMainEq}
  G\l(\tau_i\r) = \sum_j K\l(\omega_j, \tau_i\r)\rho\l(\omega_j\r)\Delta \omega.
\end{equation}
Without losing generalization, $\Delta \omega$ can be incorporated into $K$. In general, the number of the correlation function data points $G\l(\tau_i,T\r)$ is at order of $\sim O\,(10)$ accompanied with inevitable noises, however, the spectral function defined in a domain which requires around $\sim O\,(10^{3})$ data to be well constructed \cite{Asakawa:2000tr}.
One of the most intuitive way is to monitor and minimize the difference between the original $G(\tau_i)$ and the obtained $\hat{G}(\tau_i)$, where $\hat{G}$ is the correlation function written by the reconstructed SPF. However, uncertainty remains as multisolutions exist. Therefore,  additional prior information of the SPF is taken into account by different methods to help reduce the numbers of the solution, briefly explained below.

\subsection{B. Traditional Methods}

In this subsection, we will introduce some commonly used methods of analyzing the ill-posed problem.\\[-0.10in]

\underline{\emph{\textbf{Truncated Singular Value Decomposition (TSVD)}}}~ The  singular value decomposition (SVD) is a tool to decompose the singular Matrix. After introducing a truncation of SVD, the Truncated Singular Value Decomposition (TSVD) is  capable of analyzing  the ill-posed problem  in  linear equation subject ~\cite{hansen_truncated_1990,hansen_numerical_1992,chen_truncated_2017}. Since we will also implement TSVD in our RBFN method described below, we will review the TSVD method with details.

In TSVD scheme, the integral kernel is decomposed as \cite{hansen_truncated_1990}
\begin{equation}
  K\l(\omega_i, \tau_j\r)=U S V^T,\label{eq:SVDMatrixForm}
\end{equation}
where $U$ and $V$ are orthonormal matrices, and $S$ is a diagonal one. For $U=[u_1,u_2,...,u_m]\in \mathbb{R}^{m\times m}$ and $V=[v_1,v_2,...,v_n]\in \mathbb{R}^{n\times n}$, the singularly valued matrix $S=\text{diag}\l(s_1, s_2,...,s_n\r)\in \mathbb{R}^{m\times n}$ with elements ordered as:  $s_1\ge s_2 \ge ... \ge s_n$. 

In most cases, the singular values $s_i$ decrease rapidly to zero and the demanding components are reduced as a consequence. The solution of the spectral is taken the form of
\begin{equation}
  \hat{\rho}=\sum_{i=1}^n \frac{u_i^T G}{s_i}v_i ,\label{eq:SVD}
\end{equation}
where  $\hat{\rho}$ is the reconstructed spectral function deriving from the generalized inverse matrix and the reduced vector space $\{v_i\}$. As $s_{i}$ goes to zero, the weight of the corresponding basis vector grows infinitely, which mainly contributes to the high frequency part. One observes that a tiny error in the coefficients $u_i^T G$, would be amplified by the singular value $s_i$ in the denominator, known as the common instability character in ill-posed problems. And such noise is inevitable since the correlation function data $G$ comes from the expensive Monte Carlo simulation.
Indeed, as an underdetermined system, the high frequency part is actually out of control due to the complexness of its structure. To precisely predict the low-frequency part of the SPF, hence, it usually sacrifices the fine structure in the high-frequency regime but simply keep the stable parts as it has already been applied in the truncated singular value decomposition (TSVD) method. Finally, we rewrite Eq.(\ref{eq:SVD}) as
\begin{equation}
  \hat{\rho}_k=\sum_{i=1}^k \frac{u_i^T G}{s_i}v_i ,\label{eq:TSVD}
\end{equation}
where the truncating parameter $k$ is chosen through analyzing the ratio between signal and noise. It's known that the oscillations tend to increase as $k$ increases, thus then the regularized solution $\hat{\rho}_k$ behaves smoother than the original $\hat{\rho}$ as expected.

The convergence of this method has been examined in~\cite{hansen_truncated_1990} by studying the relation between $\beta_i$ and  ${s_i}$, where
\begin{equation}
  u_i^T G=\beta_i=s_i^\alpha.
  \label{eq:alpha}
\end{equation}
Here the nonnegative real constant $\alpha$ represents the decay rates of  $\beta_i$ with respect to {$s_i$}. For $\alpha > 1$, $\beta_i$ decays faster than $s_i$, the regularized solution  ${\hat{\rho}_k}$  converges to the true solution and a larger $\alpha$ leads to a better approximation. While taking noise of $G$ into consideration, a faster decay of $\beta_i$ is required to smooth it. This is called the discrete Picard condition (DPC), which is employed as a verification for TSVD and also other similar regularization methods.

The TSVD method is easy to understand and implement. However, the truncated parameter $k$  is a rather arbitrary number. Plus, this integer value is hard to deal with since the optimized solution will change from one to another uninterruptedly. Besides, it is technically difficult to  introduce the prior physical knowledge such as positivity, commonly believed asymptotic behavior and so on into the solution. \\[-0.10in]

\underline{\emph{\textbf{Tikhonov regularization}}}~ One of available improvements of TSVD approach is to employ Tikhonov regularization via introducing a continuous parameter $\lambda$ in formula, as \cite{groetsch1984theory,Dudal:2013yva}
\begin{equation}
  || G-K\rho ||^2 + \lambda || \rho||^2.
\end{equation}
The solution turns out to be:
\begin{equation}
  \rho=(K^\top K+\lambda I)^{-1}K^\top G.
\end{equation}
Plugging into the conventional SVD routine, the solution transforms to
\begin{equation}
  \rho = \sum_i \frac{s_i^2}{s_i^2+\lambda}\frac{u_i^T G}{s_i}v_i.
\end{equation}
The regularization parameter $\lambda$ modifies the converging condition especially in the region of $s_i^2\lessapprox\lambda$. Unfortunately, due to the exponential decreasing of singular values in our mocking system, the Tikhonov method does not improve much in convergence compared to TSVD scheme. \\[-0.10in]

\underline{\emph{\textbf{Maximum Entropy Method (MEM)}}}~ Another popular strategy for SPF reconstruction is Maximum Entropy Method (MEM). It is built on Bayes's theorem and mainly described by two terms. One of them, $\chi^2$, is parameterized as the Gaussian likelihood distribution  based on the central limit theorem.
Another one, named as Shannon-Jaynes entropy $S$,
serves as a regulator to restrain the deviation of the reconstructed $\rho(\omega)$ against the default model of $m(\omega)$. The Bayes' theorem is in the form of \cite{Asakawa:2000tr,Ikeda:2016czj,Gao:2014rqa,Gao:2016jka},
\begin{equation}
  P\l(\rho|\,D,I\r)\propto P\l(D|\,\rho,I\r)P\l(\rho|\, I\r),
\end{equation}
where $D$ stands for the data points of correlation functions and $I$ represents the prior knowledge about the SPF. The likelihood function reads
\begin{equation}
 P\l(D|\,\rho,I\r) \propto e^{-L}, \ \ \ \ \ \ \  L=\sum_i (D_i - \hat{D}_i)^2/\sigma_i^2,
\end{equation} \\[-0.20in]
Here $\hat{D}_i$ is the spectral representation using Eq.(\ref{eq:maineq}), $\sigma_i^2$ is the covariant matrix with all off-diagonal elements zero. This part is nothing but the standard $\chi^2$ fitting, describing the errors between the reconstructed spectral function and the original data points. Only with this part the solutions cannot be settled, and hence, the entropy $S$  is plugged in the prior probability $P\l(\rho|\,I\r)$ to encode the knowledge of SPF, which reads as \cite{Asakawa:2000tr},
\begin{equation}
 P\l(\rho|\,I\r)\propto e^{\lambda S} ,
\end{equation}\\[-0.30in]
with\\[-0.20in]
\begin{equation}
  S=\int d\omega \left(\rho(\omega)-m(\omega)-\rho(\omega)\text{log}\l[\frac{\rho(\omega)}{m(\omega)}\r]\,\right).
\end{equation}
Again, $m(\omega)$ is the smooth ansatz of SPF, and $\lambda$ is a real and positive parameter.
The entropy term exponentially increases while the spectral function deviates from the default model, and then, the oscillated solution would be diminished rapidly. It is proved that the solution is becoming unique after introducing entropy $S$. \newline

We emphasize here that the traditional methods often assume the solution is the minimally-possible-oscillated function. This smoothness assumption  serves as an additional constraint in constructing SPF, adopted in many methods~\cite{hansen_truncated_1990,groetsch1984theory,Asakawa:2000tr}. To check the robustness and plausibility of such prior estimation in the traditional methods, on the contrary, we here apply a new approach free of this assumption based on the radial basis function networks.
\section{III. Radial Basis Function Networks}
In this section, we will explain the constructed Radial Basis Function Networks, which are used to extract the spectral functions from the correlation data in this paper.
Radial Basis Function refers to a function taking form of $\phi(||\boldsymbol{x}-\boldsymbol{m}||)$, where $||\boldsymbol{\cdot}||$ is the Euclidean norm and measures the distance of input $\boldsymbol{x}$ and some fixed point $\boldsymbol{m}$. Radial basis function network is a simple neural network that uses RBFs as activation functions, where the output is a weighted linear combination of RBFs of the inputs. They have been widely applied to interpolate scattered data and approximate multivariate functions~\cite{micchelli_interpolation_1986,broomhead1988radial,chen1991orthogonal,park1993approximation}, solving equations~\cite{dyn1983iterative,kansa_multiquadricsscattered_1990,golbabai_numerical_2006}, etc.

In principle, an arbitrary function $\rho\l(\omega\r)$ (including the spectral function) can be approximately described by a linear combination of radial basis functions:
\begin{equation}
  \rho\l(\omega\r)=\sum_{j=1}^N w_{j}\phi\l(\omega-m_j\r),\label{eq:linear_summation}
\end{equation}
where $\phi$ are the active RBFs, $w_{j}$ is the weight to be determined by optimization, while the centers of a radial basis function $m_j$ should be determined artificially before optimization. A number of functions can be applied as the RBF, such as radial distance, thin-plate spline, compact support, etc. In this paper, we employ two types of RBFs, Gaussian and MQ, respectively. They are written as:
\begin{equation}\label{eq:RBFs}
\begin{aligned}
  \text{Gaussian:}\ \ \ \phi(r)&=e^{-\frac{r^2}{2a^2}}, \\
  \text{MQ:}\ \ \ \phi(r)&=(r^2+a^2)^{\frac{1}{2}}.
\end{aligned}
\end{equation}
where the adjustable value of $a$ in Gaussian or MQ is known as the shape parameter, which is essential for the regularization. In general, the shape parameter $a$ is associated with the distance between adjacent centers and controls the smoothness of the interpolating function. With SPFs  described as a combination of certain Gaussian or MQ basis functions, the related solution can be more easily obtained. Note that the basis  could  be  altered to some other  functions to suit the specific requirement of  certain problems.

\begin{figure}[htbp]
  \centering
  \includegraphics[width=3.5cm]{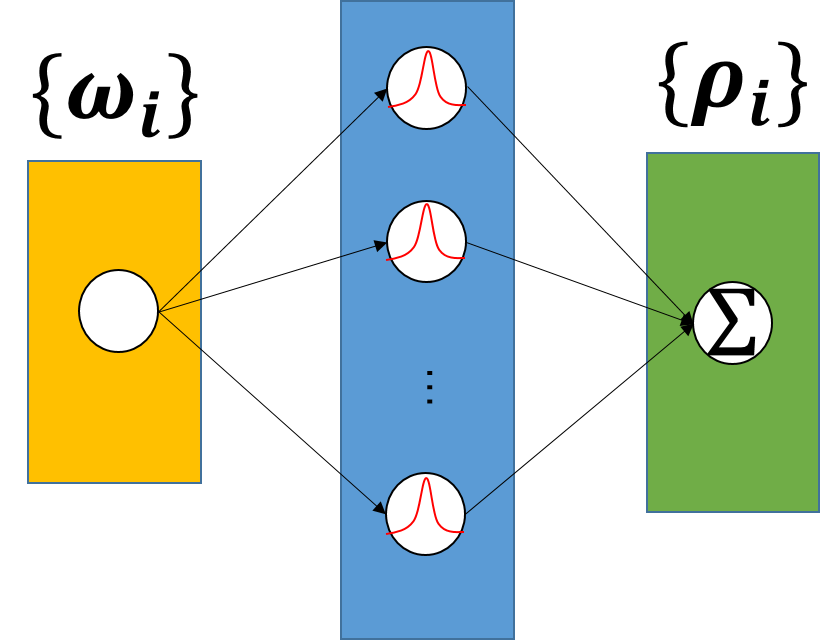}
  \caption{Illustration of the structure of RBF networks.}
  \label{fig:rbf_structure}
\end{figure}

Radial basis function networks is a three layers feed-forward neural network with the active RBFs built in the hidden layer (as illustrated by Fig.(\ref{fig:rbf_structure})). In more detail, the RBFN transform Eq.(\ref{eq:linear_summation}) into its matrix form:
\begin{equation}
  \l[\rho\r]=\l[\Phi\r]\l[W\r], \label{eq:matform}
\end{equation}
and then substitute it into the integral equation Eq.(\ref{eq:maineq}).  $\l[\Phi\r]_{ij}=\phi\l(||\omega_i-m_j||\r)$ is an $N\times M$ matrix, called the interpolation matrix. Here, we discrete the spectral function as $\rho(\omega_i)$ and set $m_i=\omega_i,\ i=1...N$,  with $M=N=500$ \footnote{We found that the matrix with $M=N=500$ is sufficient to construct the several spectral function presented in this paper.  Larger matrix does not improve the results.}. In this case every data point of the target function is directly assigned with a RBF. Consequently, $\Phi=\Phi^T$.  The shape parameter $a$ is set according to prior assumptions of the SPF, such as positivity, etc. Please refer to the appendix for details.

\begin{figure*}
  \begin{center}
    \includegraphics[width=0.8\textwidth]{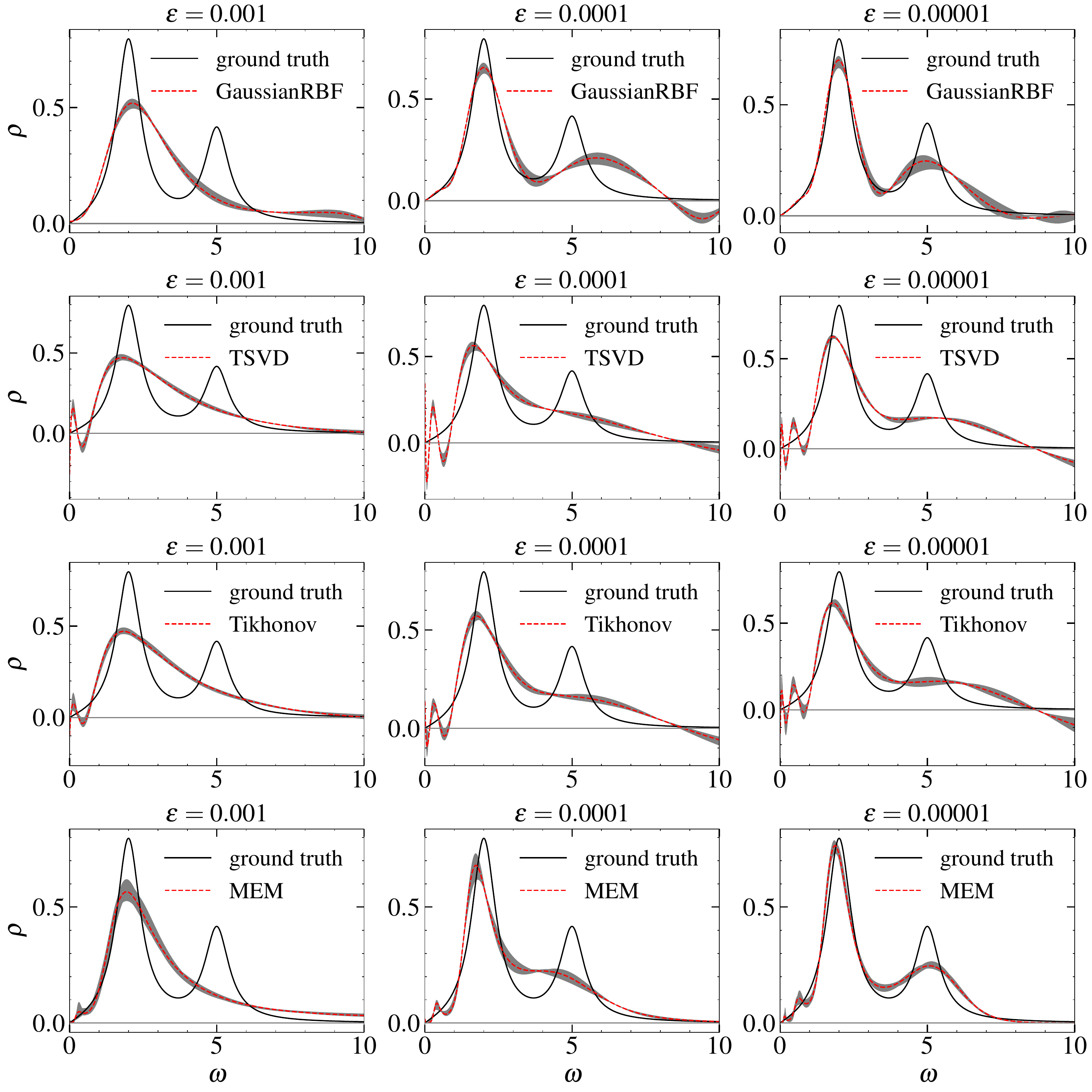}
    \caption{The predicted spectral functions from RBFN, TSVD, Tikhonov and MEM, using the correlation data generated by the mock SPF Eq.(\ref{eq:Mock SPF}) with parameter set I (para-I). For left to right panels,  different Gaussian noises are added to the correlation data with $\epsilon = 0.001$,  $0.0001$ and  $0.00001$.}
    \label{fig:RBFNoise}
  \end{center}
  \end{figure*}

After applying the RBFN representation, Eq.(\ref{eq:MatFormOfMainEq}) becomes:
\begin{equation}\label{eq:RBFMatForm}
  G_i=\sum_{j=1}^{M} \sum_{k=1}^N K_{ij}\Phi_{jk}w_k\equiv \sum_{k=1}^{M} \tilde{K}_{ik}w_k, \ \ \ i=1...\widehat{N}
\end{equation}
where $\tilde{K}$ is a $\widehat{N}\times M $ matrix, which is irreversible.  In the following calculation in Sec.IV, we generate $\widehat{N} =30$ data points for the correlation function $G_i$, using some specific mock SPF. To obtain $w_j$, we further implement the TSVD method~\footnote{$w_i$ can also be obtained by machine learning using the
gradient descent algorithm. While for the simple network structure shown by Fig.(\ref{fig:rbf_structure}),  it costs more calculation time and the results may not be stable, compared with the TSVD method used here.} described above:
\begin{equation}
  \boldsymbol{w}=\sum_{i=1}^k \frac{\tilde{u}_i^T G}{\tilde{s}_i}\tilde{\boldsymbol{v}}_i ,\label{eq:TSVD_RBF}
\end{equation}
where  $\tilde{u}_i, \tilde{s}_i, \tilde{\boldsymbol{v}}_i$ are decomposed according to Eq.(\ref{eq:SVDMatrixForm}) from $\tilde{K}_{ik}$ defined in Eq.(\ref{eq:RBFMatForm}). The summation cutoff $k$ is set by hand and equal to $10$ in the following calculation, which will be explained in the appendix.

Compared with the traditional Tikhonov method and Maximum Entropy Method mentioned above, which treat the SPF as discrete data points, our RBFN scheme is genuinely meshless and mathematically straightforward. Besides, compared with other neural networks methods, such as supervised learning approach~\cite{yoon2018analytic, fournier2020artificial, Kades:2019wtd}, it is rapidly trained and free from the bias and overfitting problem.

\section{IV. RESULTS}

\begin{figure}[t]
  \begin{center}
    \includegraphics[width=1.0\textwidth]{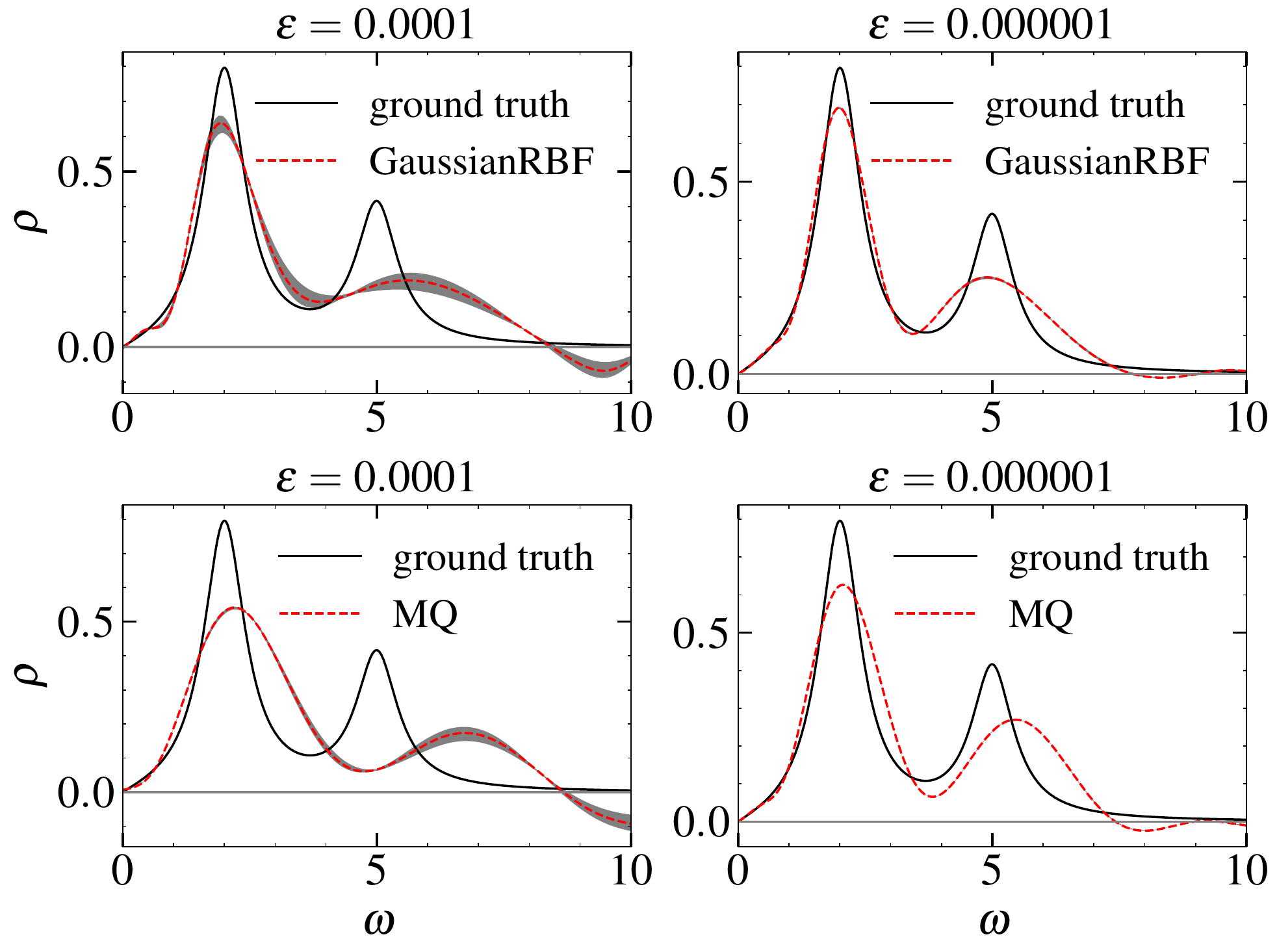}  
    \caption{The predicted spectral functions from our RBFN with Gaussian and MQ RBF. The correlation data are generated with the same mock SPF as used in Fig.~(\ref{fig:RBFNoise}). }
    \label{fig:RBFComp}
  \end{center}
\end{figure}
\begin{figure}[t]
  \begin{center}
    \includegraphics[width=1.0\textwidth]{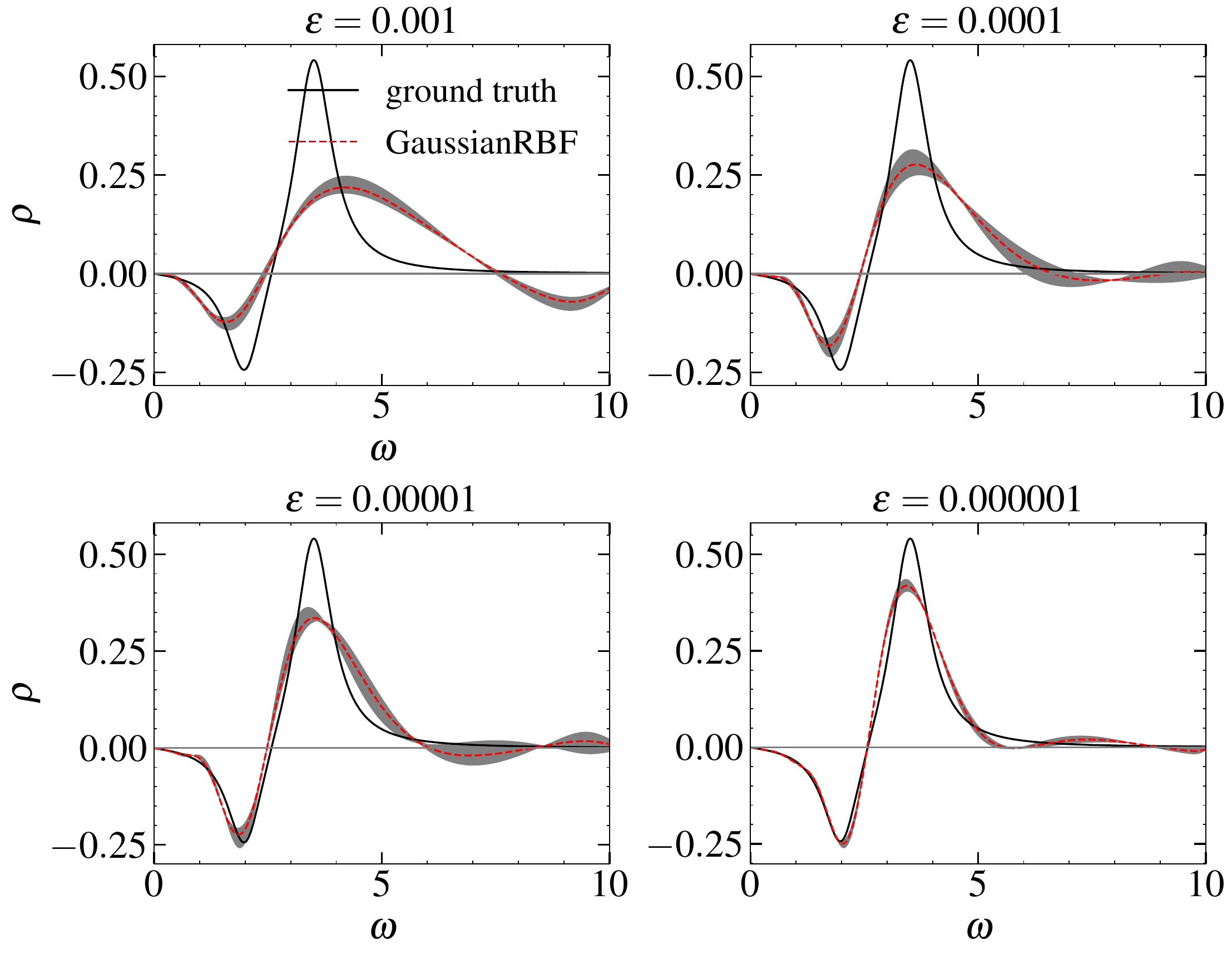} 
    \caption{The predicted spectral functions from our RBFN (with Gaussian RBF), using the correlation data generated by mock SPF Eq.(\ref{eq:Mock SPF}) with para-II. Four different Gaussian noises $\epsilon=0.001$, $\epsilon=0.0001$, $\epsilon=0.00001$ and $\epsilon=0.000001$ are added to the correlation data. }
    \label{fig:RBFNeg}
  \end{center}
\end{figure}


In this section, we will implement RBFN method for the correlators
 generated by two types of  mock SPF, associated with  propagators~\cite{boon1991molecular,forster2018hydrodynamic,Ding:2010ga} and  stress energy tensor~\cite{Qin:2014dqa}, respectively.
The first type can be applied to describe the quasi-particles, which are especially useful for studying the phase transition from hadron phase to QGP phase. The second type of SPF contains a resonance peak at low frequency and a continuum part at large frequency, which can be used to extract the transport coefficients, such as the diffusion coefficient of heavy quarks.  \\

We firstly construct the propagators  by mixing several Breit-Wigner distributions:
\begin{equation}\label{eq:Mock SPF}
  \rho_{Mock}(\omega)=\sum_{i}\rho_{BW}(A_i,\Gamma_i, M_i, \omega)
\end{equation}\\[-0.20in]
with \\[-0.30in]
\begin{equation}
 \rho_{BW}(A_i,\Gamma_i, M_i, \omega)=\frac{4A_i\Gamma_i \omega}{\l(M_i^2+\Gamma_i^2-\omega^2\r)^2+4\Gamma_i^2 \omega^2},
\end{equation}
where $A_i$ is the normalization parameter, $M_i$ denotes the mass of the particle, carrying the location of the peak, and $\Gamma_i$ is the width. For the calculations in Fig.~(\ref{fig:RBFNoise}) and Fig.~(\ref{fig:RBFComp}),  the mock SPFs are constructed through combination of two Breit-Wigner peaks using $A_1=0.8, M_1=2, \Gamma_1=0.5; A_2=1, M_2=5, \Gamma_2=0.5$ (para-I).  For the calculations in Fig.~(\ref{fig:RBFNeg}), the parameters are set to $A_1=-0.3, M_1=2, \Gamma_1=0.5; A_2=1, M_2=3.5, \Gamma_2=0.5$ (para-II), which make the first peak of the mock SPF become negative. With the constructed mock SPF, we then generate the Euclidean correlation functions $G(\tau_i)$ according to Eq.(\ref{eq:maineq}) ($T=0$), together with noise added to the mock data:
\begin{equation}\label{eq:G_noise}
  G_{\text{noise}}(\tau_i) = G(\tau_i) + \text{noise}.
\end{equation}
Here we generate $30$  discrete correlation data  evenly distributed between $\tau = 0$ and $10$. The noises are randomly generated by the Gaussian distribution similar to \cite{Asakawa:2000tr} where the mean are set to 0 and the width $\epsilon$  is a adjusted parameter that corresponds to different noise levels.

Fig.(\ref{fig:RBFNoise}) compares the predicted spectral functions from RBFN, TSVD, Tikhonov and MEM using the correlation data described above.
For each prediction, the solution is obtained by averaging $10$ results using the sampled correlation data with random Gaussian noises. The shaded area denotes the uncertainty for each model prediction.  Although the Tikhonov method shows a tiny improvement with the infrared parameter $\lambda$, both Tikhonov and  TSVD  share the same oscillation behavior at the low frequency part. In contrast, RBFN and MEM provide better predictions at the low frequency part.  The results from RBFN do not oscillate and almost reproduce the first peak of the mock SPF for the correlation data with smaller noise $\epsilon = 0.00001$. This is especially important for such a task of extracting the transport coefficients from the Kubo relation described by Eq.(\ref{eqn:eta}).  Our RBFN is the only method that  reduces the oscillation in the low frequency area, compared with the other three commonly used approaches.

At high frequencies, all four studies fail to reconstruct the second peak of the mock SPF from the discrete Euclidean correlators  with larger gaussian noises  $\epsilon=0.001$.  After reducing the gaussian noises  to $\epsilon=0.00001$, both RBFN and MEM can approximately reproduce the second peak of the mock SPF. In contrast, Tikhonov and TSVD methods fail to do so.  Although  the behavior of RBFN results at high frequency is somewhat spoiled, it might be improved by imposing the positivity condition, which we would like to leave to further investigation.

In Fig.(\ref{fig:RBFComp}), we compare the predicted  spectral functions from our RBFN method using two types of RBFs called Gaussian and MQ  described by Eq.(\ref{eq:RBFs}). The correlation data are generated with the same mock SPF as used in Fig.(\ref{fig:RBFNoise}), using different levels of Gaussian noises. These two types of RBFN predict similar results, while the one with Gaussian RBF is more powerful to reconstruct the locating of the peaks for the spectral function. It also almost describes the width of the first peak for the mock SPF.

In Fig.(\ref{fig:RBFNeg}), we test our RBFN (with the Gaussian RBF), using the correlation data generated from the two-peak mock SPF with the first peak turning negative (para-II as described above). It turns out that our RBFN works consistently well in constructing such spectral functions  at the low frequency limit. The location of the peak has been well represented within a range of allowable errors, especially for the correlation data with smaller Gaussian noises.
Again, the  behaviors of the constructed spectral function  are somewhat spoiled  at high frequency regime for the correlation data with larger noises,  which
are similar to the cases in Fig.(\ref{fig:RBFNoise}) and Fig.(\ref{fig:RBFComp}).\\

\begin{figure}[t]
\begin{center}
    \includegraphics[width=0.9\textwidth]{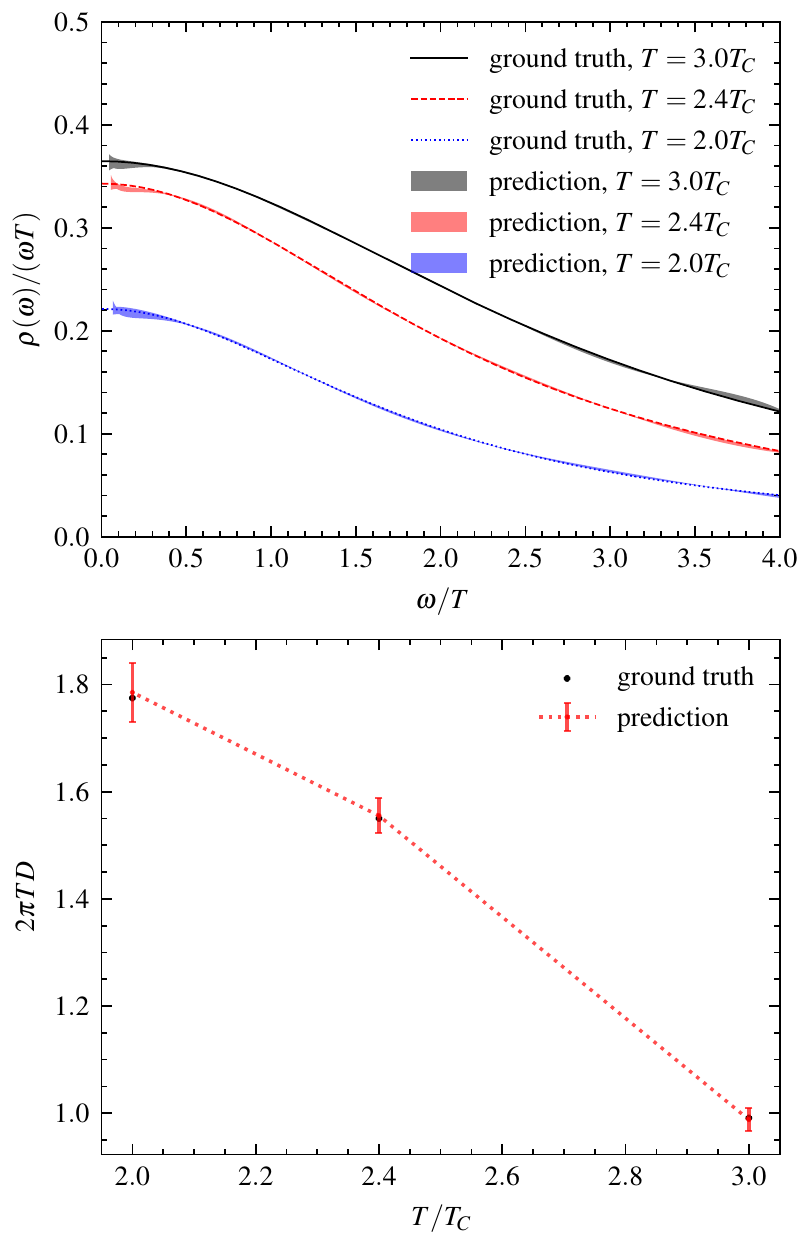}
       \caption{The spectral functions (upper panel) and corresponding diffusion coefficients $D$ (lower panel)  at different temperatures, predicted by our RBFN method with Gaussian RBF. }    \label{fig:low_freq}
       \end{center}
\end{figure}

Now we test our RBFN using another type of spectral function associated with the heavy quark diffusion in the QGP~\cite{Qin:2014dqa}:
\begin{equation}\label{eq:transportSpectral}
\begin{split}
  \rho_V(\omega)=&\frac{6\chi_{00}T}{M_0}\frac{\omega\eta_D}{\omega^2 + \eta_D^2} + \frac{3}{2\pi}\Theta(\omega^2-4M_0^2)\\
  &\times \omega^2 \text{tanh}(\omega/4T)\sqrt{1-4M^2_0/\omega^2}\\
  &\times [1+4M^2_0/\omega^2].
\end{split}
\end{equation}
where $\chi_{00}$ is the quark number susceptibility, $\eta_D=\frac{T}{M_0 D}$ is the drag coefficient and $M_0$ is the thermal quark mass. Following~\cite{Qin:2014dqa}, we set $M_0$ to be $1$ GeV and $\eta_D\approx \frac{0.9T^2}{M_0}$. The transport peak in the infrared part determines the transport coefficients, while the continuous part in the perturbative region can be calculated perturbatively. Therefore,  we use the analytical form above $1$ GeV and apply our RBFN method with Gaussian RBF to determine the spectral function below.

Using Eq.(\ref{eq:RBFMatForm}) and the above spectral function, again, we generate $30$ discrete correlation data that are evenly distributed between $\tau = 0$ and $10$. To testify the stability, we also put small random truncated Gaussian noises with width $1$ and truncation between $-5$ MeV and $5$ MeV to $M_0$ to mimic the possible deviation between the numerical data and the analytical form of the continuous part.

The upper panel of Fig.(\ref{fig:low_freq}) shows our RBFN's predictions for the low-frequency parts of the spectral function.  In general, our method gives a stable and smooth spectral in the infrared regime that nicely fits the ground truth described by Eq.(\ref{eq:transportSpectral}) for various temperatures.  Due to numerical instability closing zero,  we have applied a cut-off for $\omega$ at $\sim 10^{-2}$,  which corresponds to $\omega/T\sim 10^{-1}$ in the figure. We have added noises to $M_0$ in the first place, and that is where the uncertainties mainly come from. The quark diffusion coefficient $D$  can be obtained from the diffusive Kubo formula $ D=\frac{1}{6\chi_{00}}\lim_{\omega\rightarrow 0}\sum^3_{i=0}\frac{\rho^{ii}_V(\omega)}{\omega}$. As shown by the lower panel of Fig.(\ref{fig:low_freq}), our prediction is consistent with the analytical result obtained from Eq.(\ref{eq:transportSpectral}).


\section{V. Summary}

A direct computation of spectral function is extremely difficult in QCD because of confinement. The reconstruction from numerical data computed in Euclidean space is an alternative way.  As a typical ill-posed problem,
several algorithms have been applied in reconstructing such as  TSVD, Tikhonov method, and MEM.
In this paper, we  developed a machine learning technique for the extracting process. Unlike other commonly used discretized schemes, a continuous neural networks representation based on  the radial basis functions has been adopted.

We first applied it for building the spectral of propagators. Our method shows a consistent reconstruction for different types of propagators which are parametrized with several Breit-Wigner distributions.  In detail, the location and width of peak for the propagator with positive definite and negative spectral can be both well reproduced within  allowable errors. This is especially useful considering the confinement-deconfinement phase transition, since it has been widely conjectured that the spectral of confined states have a negative part in contrast to that of the deconfined states which are positively described.

We also compare two types of RBF in our analysis, Gaussian and MQ ones. It is found that both obtain the SPF efficiently, but Gaussian RBF leads to a better fitting for the location, as well as the width of the peak. Therefore, the Gaussian RBF generally better suits the problem considered here, the reconstruction of SPF.

Moreover, checked by a known formulation of the spectral of energy stress tensor, our approach shows a significant performance in the low frequency region
and presents a stable result for the diffusion coefficients comparing to other traditional ways. This benefit is with great importance for the future studies of transport coefficients. Its applications in the low energy regime, combining with the non-perturbative QCD calculations, will be explored elsewhere.

\section{VI. Appendix}
\begin{figure}[t]
  \includegraphics[width=1.0\textwidth]{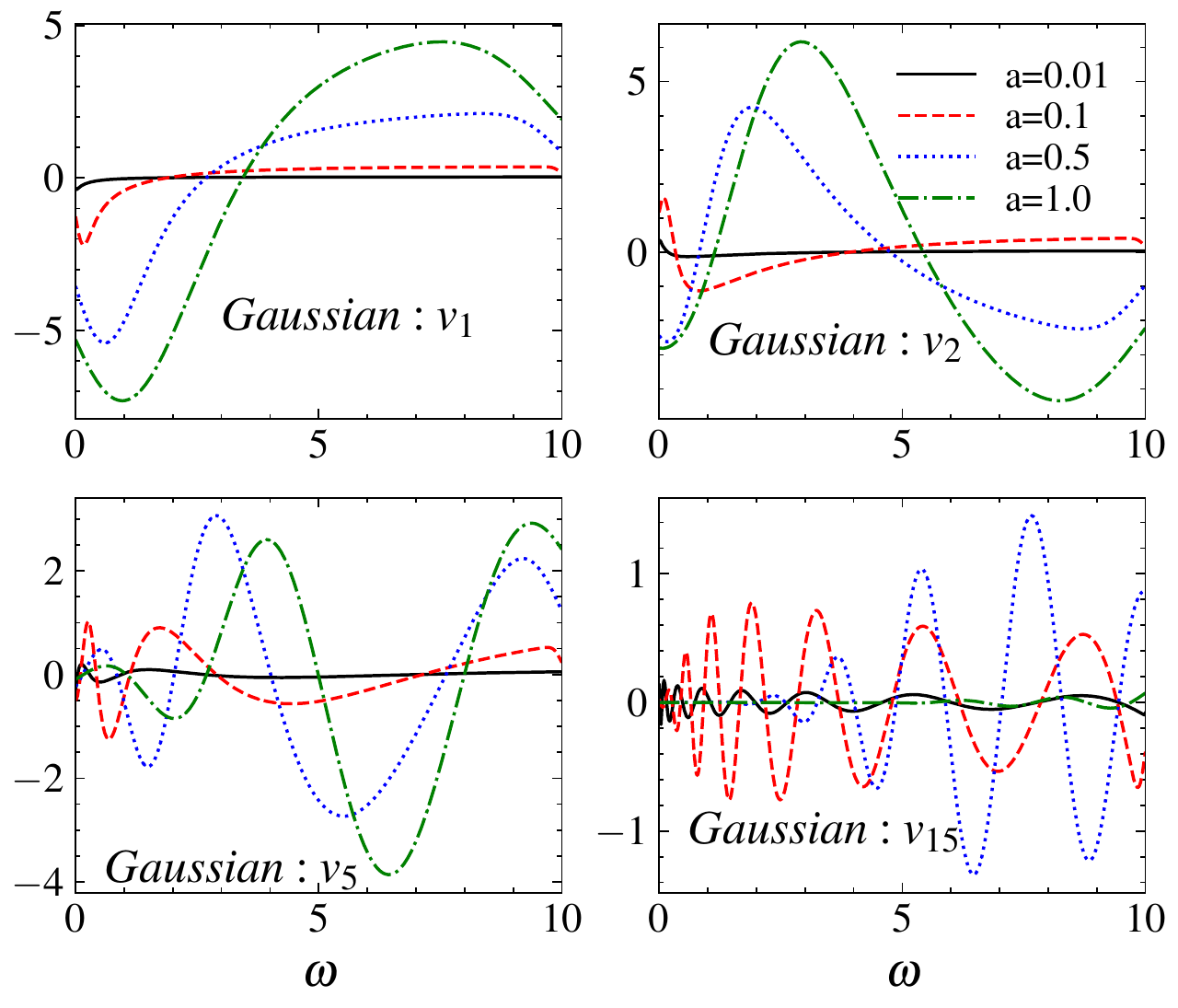} 
  \caption{Several basis vectors for different shape parameters  $a$ in Gaussian RBFN.}
  \label{fig:RBFSubspace_Gaussian}
\end{figure}

\begin{figure}[t]
  \includegraphics[width=1.0\textwidth]{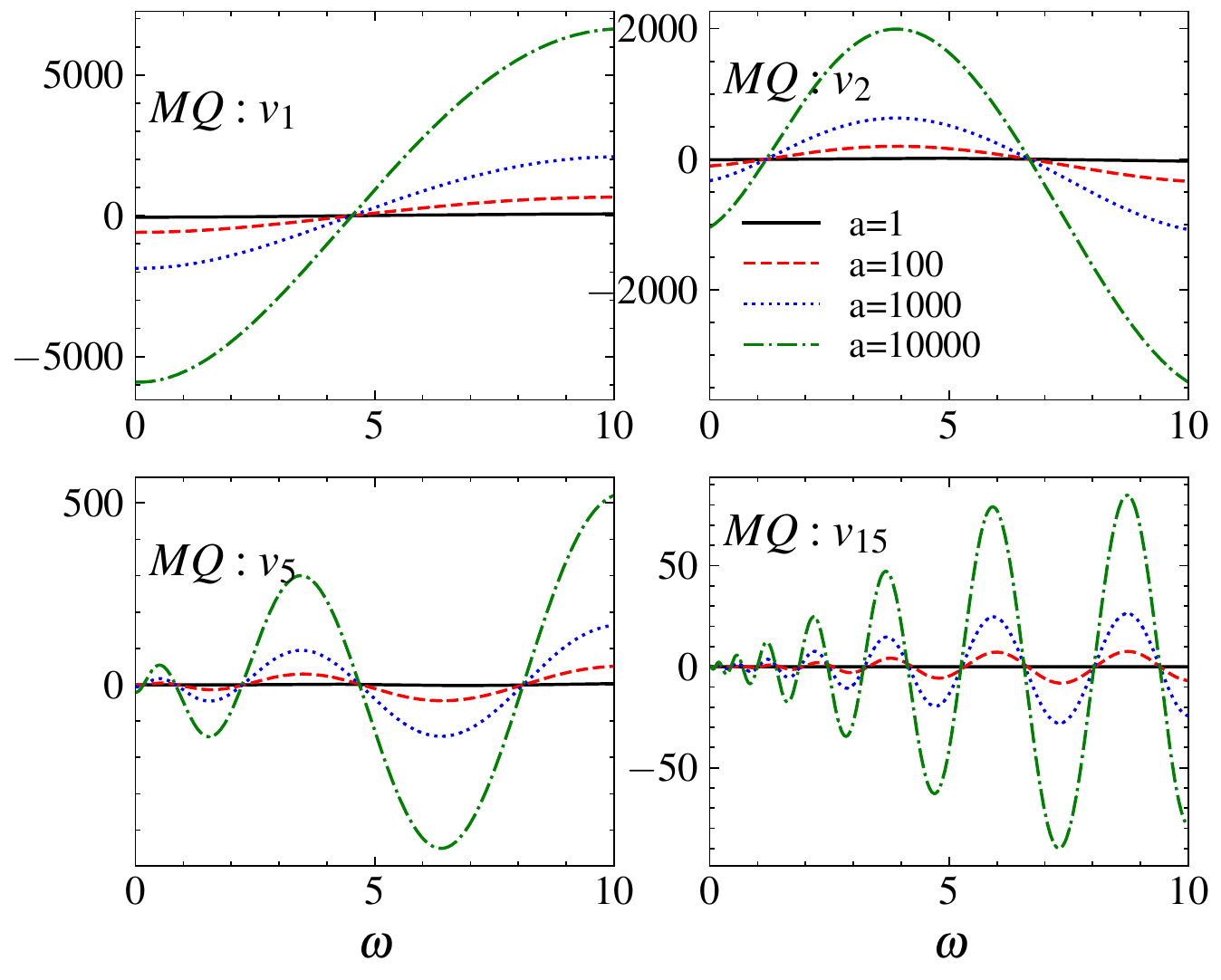} 
  \caption{Several basis vectors for different shape parameters $a$ in MQ RBFN.}
  \label{fig:RBFSubspace_MQ}
\end{figure}

\begin{figure}[t]
  \includegraphics[width=1.0\textwidth]{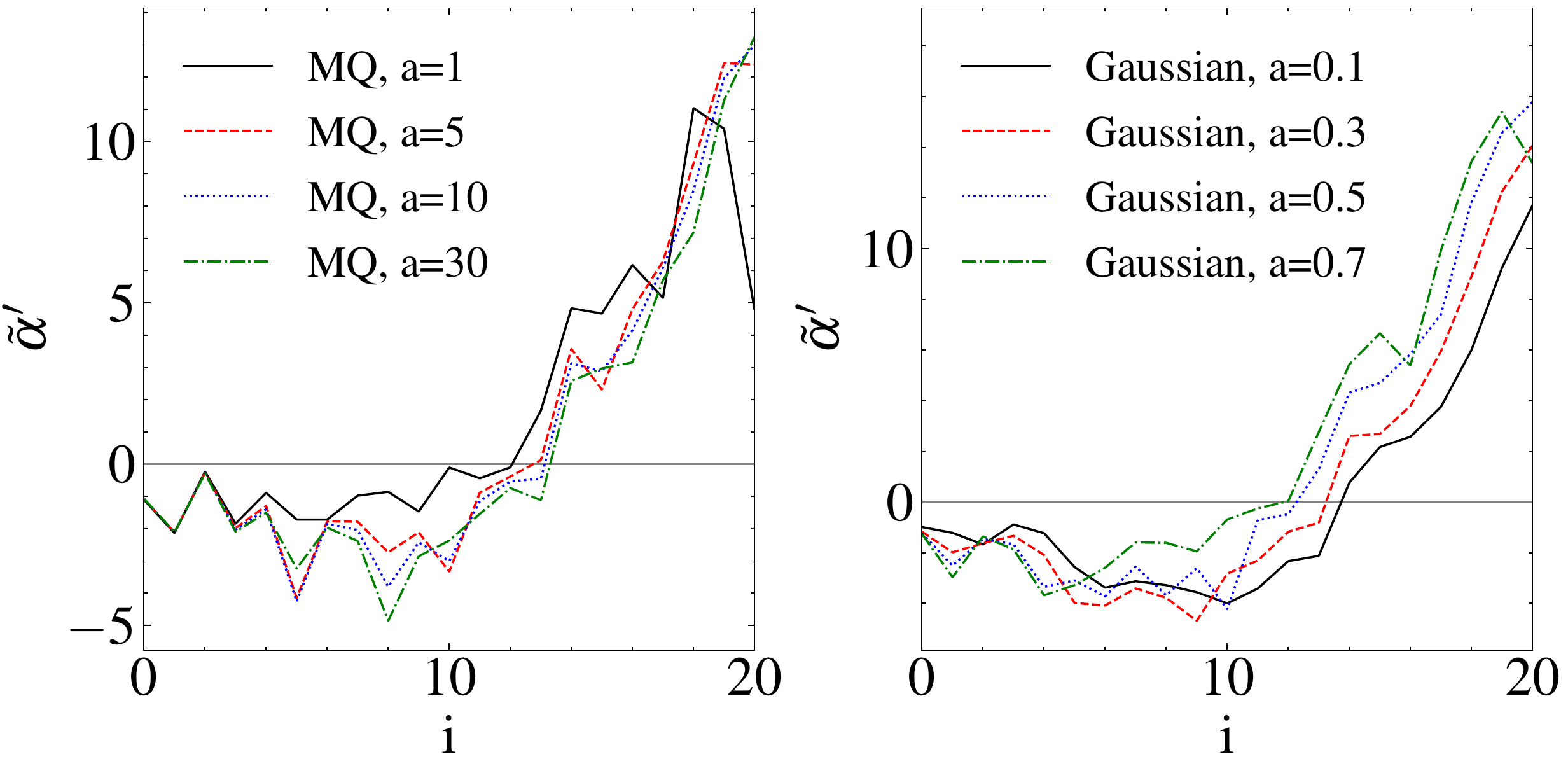}  
  \caption{The $\tilde{\alpha}'$s of the MQ with different shape parameters $a$ (left) and the $\tilde{\alpha}'$s of the Gaussians with different shape parameters $a$ (right). $i$ stands for the index for a basis vector. When the index $i$ growing, $\tilde{\alpha}'$ become positive rapidly.}
  \label{fig:decay_Gaussian_MQ}
\end{figure}

Fig.(\ref{fig:RBFSubspace_Gaussian}) and Fig.(\ref{fig:RBFSubspace_MQ}) show the basis vectors $\boldsymbol{v_i}$
for different shape parameters  $a$ in Gaussian RBFN and MQ RBFN, obtained by the TSVD method  in Eq.(\ref{eq:TSVD_RBF}).
As the index $i$ increases, $v_i$ strongly oscillates  with $\omega$.  For MQs RBFN, $a$ modifies the magnitude of $v_i$, but does not change their shape. For Gaussian RBFN, $a$ affects both the shape and magnitude of $v_i$. In the calculations shown in Sec.IV, $a$ is set to be around $0.4$, which improves the positivity of the predicted SPF. 
{Without such a physical constraint, the spectral function can still be extracted but with slightly larger errors.}

Except the shape parameter, a truncation of the basis space associated with the index $\alpha$ is also required, similar to the TSVD method. Nevertheless, the subspace can be simultaneously determined in the RBFN method since the effect of noise in the data is presented by $\alpha$ and in turn, this decaying index will help to choose the "good" components according to the discrete Picard condition. We denote the criterion by $\tilde{\alpha}_i=\log\l(\beta_i/s_i\r)$. If $\tilde{\alpha}_i>0$, the respective $i$-th basis is strongly affected by the noise, while for $\tilde{\alpha}_i\leq 0$, it indicates the corresponding vectors are stable against the noise. The behavior of $\{\tilde{\alpha}_i\}$ is shown in Fig.(\ref{fig:decay_Gaussian_MQ}). It is worth mentioning that the "good" components which selected by the RBF method are not always the minimally-oscillated functions in contrast to the conventional ways. We also expect that the formed truncation scheme, controlling by the noise, will serve as a guideline in the TSVD approach.

\hspace{2cm}
\begin{acknowledgments}
We thank M. Asakawa for helpful discussions.  The work is supported by the NSFC under grant Nos.~12075007 and ~11675004. F.~Gao is supported by the Alexander von Humboldt foundation. We also gratefully acknowledge the extensive computing resources provided by the Super-computing Center of Chinese Academy of Science (SCCAS), Tianhe-1A from the National Supercomputing Center in Tianjin, China and the High-performance Computing Platform of Peking University.
\end{acknowledgments}

\enlargethispage{1cm}

\bibliography{papNotes}

\end{document}

%% file: commands.tex
\newcommand{\com}[1]       {\relax}

